\documentclass{article}

\usepackage[english]{babel}

\usepackage[letterpaper,top=2cm,bottom=2cm,left=3cm,right=3cm,marginparwidth=1.75cm]{geometry}

\usepackage{amsmath}
\usepackage{graphicx}
\usepackage[colorlinks=true, allcolors=blue]{hyperref}
\usepackage{enumitem}

\title{A Case Study on Test Case Construction with Large Language Models: Unveiling Practical Insights and Challenges}
\author{
  Roberto Francisco de Lima Junior\\
  \texttt{rflj@cesar.org.br}
  \and
  Luiz Fernando Paes de Barros Presta\\
  \texttt{lfpbp@cesar.school}
  \and
  Lucca Santos Borborema\\
  \texttt{lsb@cesar.school}
  \and
  Vanderson Nogueira da Silva\\
  \texttt{vns@cesar.org.br}
  \and
  Marcio Leal de Melo Dahia\\
  \texttt{mlmd@cesar.org.br}
  \and
  Anderson Carlos Sousa e Santos\\
  \texttt{acss@cesar.org.br}
}

\begin{document}
\maketitle

\begin{abstract}
This paper presents a detailed case study examining the application of Large Language Models (LLMs) in the construction of test cases within the context of software engineering. LLMs, characterized by their advanced natural language processing capabilities, are increasingly garnering attention as tools to automate and enhance various aspects of the software development life cycle. Leveraging a case study methodology, we systematically explore the integration of LLMs in the test case construction process, aiming to shed light on their practical efficacy, challenges encountered, and implications for software quality assurance.

The study encompasses the selection of a representative software application, the formulation of test case construction methodologies employing LLMs, and the subsequent evaluation of outcomes. Through a blend of qualitative and quantitative analyses, this study assesses the impact of LLMs on test case comprehensiveness, accuracy, and efficiency. Additionally, delves into challenges such as model interpretability and adaptation to diverse software contexts.

The findings from this case study contributes with nuanced insights into the practical utility of LLMs in the domain of test case construction, elucidating their potential benefits and limitations. By addressing real-world scenarios and complexities, this research aims to inform software practitioners and researchers alike about the tangible implications of incorporating LLMs into the software testing landscape, fostering a more comprehensive understanding of their role in optimizing the software development process.
\end{abstract}

\section{Introduction}

Software testing constitutes a critical phase in the software development life cycle, ensuring the delivery of reliable and high-quality software products. In recent years, the advent of advanced technologies, particularly Large Language Models (LLMs), has introduced novel possibilities for automating and enhancing various facets of software engineering. LLMs, exemplified by models such as OpenAI's GPT (Generative Pre-trained Transformer), exhibit remarkable natural language processing capabilities and have demonstrated applicability in diverse domains. This paper addresses the intersection of LLMs and software testing, focusing specifically on their role in the construction of test cases.

The use of LLMs in software testing holds the promise of increasing the efficiency and effectiveness of test case generation. The ability of these models to understand and generate human-like text prompts paves the way for more natural and expressive articulation of test scenarios. However, the practical implications of integrating LLMs into the nuanced process of test case construction remain under explored. This study seeks to bridge this gap by presenting a detailed case study that investigates the application of LLMs in constructing test cases for a real-world software application.

The case study methodology adopted in this research involves the selection of a representative software application, the formulation of test case construction methodologies leveraging LLMs, and a thorough evaluation of the outcomes. By examining the interplay between LLMs and the complexities inherent in software testing, this study aims to provide insights into the practical advantages, challenges, and considerations associated with their implementation. Through this exploration, the intent is to contribute substantively to the ongoing dialogue surrounding the integration of cutting-edge technologies, such as LLMs, in the pursuit of optimizing software testing practices.

\section{Related Work}

The intersection of Large Language Models (LLMs) and software engineering has witnessed growing interest in recent literature, reflecting a broader acknowledgment of the transformative potential of these models in many domains. Within the realm of software testing, where the efficacy of test case construction significantly influences software quality, the exploration of LLM applications remains a relatively nascent but promising area.

The preeminent contribution of LLMs to software engineering lies in their ability to comprehend and generate human-like text, a feature that has inspired its application in diverse natural language processing tasks. Prior studies, such as \cite{radford2018improving} showcase the prowess of LLMs, exemplified by models like GPT-3, in understanding context, generating coherent responses, and adapting to diverse language inputs. This linguistic versatility positions LLMs as potentially powerful tools for expressing and formulating test cases in a more natural, context-aware manner.

Taking advantage of LLMs, many Generative AI models have been developed to implement diverse use cases, such as generating documentation for code \cite{feng2020codebert}, auto-completing code  \cite{chen2021evaluating, kim2021code}, generating unit tests \cite{tufano2021unit} , and finding duplicated code \cite{guo2021graphcodebert}.

Models trained on extensive code data sets can effectively manage multiple use cases concurrently, as exemplified by PLBART \cite{ahmad-etal-2021-unified} , CodeBERT \cite{feng2020codebert} , and GraphCodeBERT \cite{guo2021graphcodebert}. Notably, OpenAI introduced Codex \cite{chen2021evaluating} , a GPT-based model trained on GitHub code, powering their CoPilot \cite{githubcopilot} product. This model excels in auto-completing code across various programming languages, such as Python, TypeScript, Go, and Ruby, in addition to generating code from natural language.

In the specific context of software testing, recent research by \cite{wang2023software} summarized the use of LLMs in a diverse set of applications, integrating them with traditional methods such as differential and mutation testing. Despite showing promise, LLMs are recognized as complementary rather than exclusive solutions for software testing. The research also combined LLMs with metamorphic testing, generating test cases based on expected input-output relationships, and explored their role in model-based testing, enhancing model accuracy through natural language descriptions and adapting to system changes. 


However, existing literature lacks comprehensive insights into the practical implications, challenges, and considerations associated with integrating LLMs into the intricate task of test case construction for real-world software applications.



\section{Approach}

This section outlines our approach, consisting of the semi-automated test cases constructions and the evaluation framework for it. First it was explored the integration of LLMs with software specifications for generating test cases and later elucidates our method for qualitatively assess the effectiveness of the generated test cases.

\subsection{Test cases construction}

To generate test cases from the model, prompt engineering was employed. The prompt, which serves as the input for the LLM, comprises a set of instructions and necessary data for task completion. The quality of the model's response improves with clearer and more contextual descriptions provided in the prompt. Another effective strategy for achieving positive outcomes involves allowing the model to engage in reasoning by decomposing the task into intermediate steps~\cite{wei2023chainofthought}.

\begin{figure}[!ht]
\centering
\includegraphics[width=\linewidth]{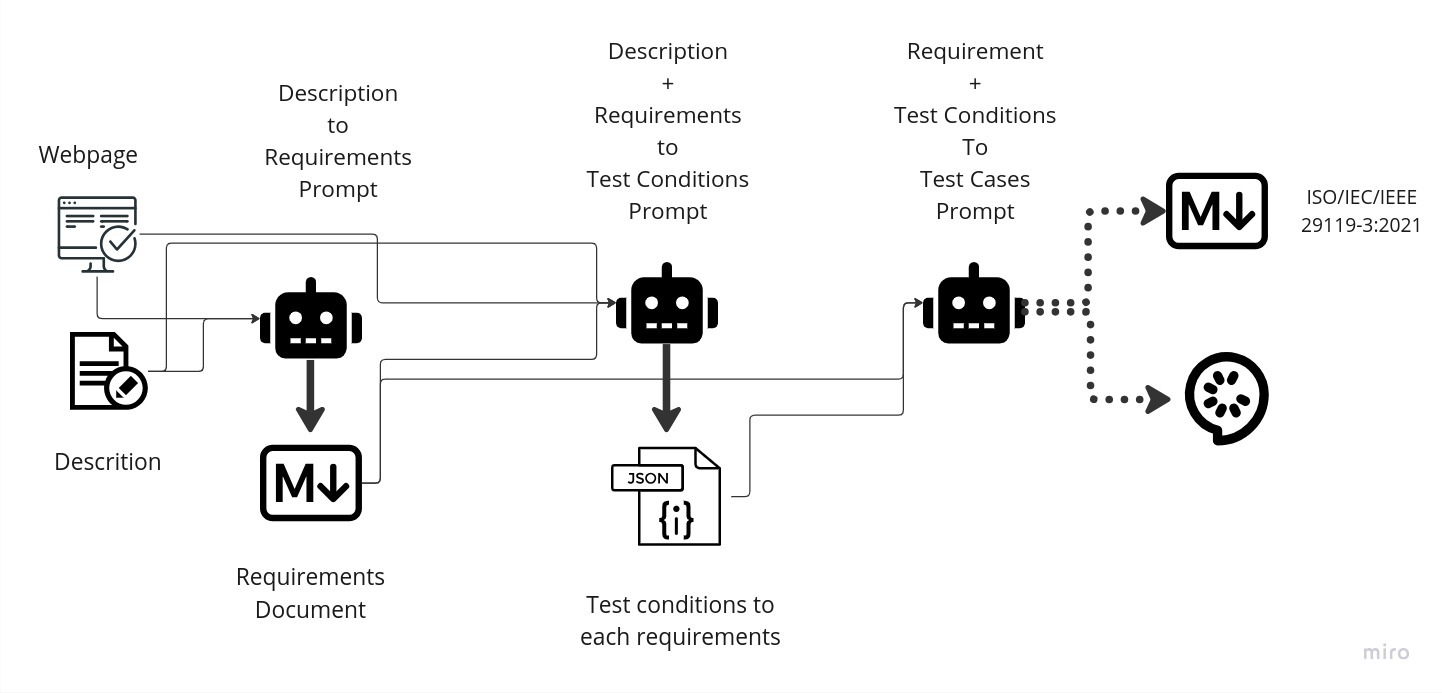}
\caption{\label{fig:test_cases_with_ai} The workflow for building test cases using a Large Language Model (represented as the robot)}
\end{figure}

Considering this, an interactive method was designed to extract test cases from the model. Figure~\ref{fig:test_cases_with_ai} provides an overview of this process. Initially, a manual application description was created to supply the essential information. To facilitate this, a predefined template was created, aiding the engineer in providing necessary details and relevant data quicker. This template was empirically extensively tested to check which kind of information would be more relevant to GPT understand the software context better and with less data. The chosen questions are presented in subsection \textbf{3.1.1}. Following that, the model is guided to generate a requirements document. This document is then used in along with the previously description to formulate test conditions. These test conditions are subsequently combined with the requirements to generate test cases in the desired format.

The OpenAI API was used with the GPT-3.5 Turbo model \cite{openai2023} together with the Python framework LangChain \cite{langchain} in order to manage the prompts, insert the context, memory, and handle interactivity. The following subsections details the step by step of this procedure.

\subsubsection{Application Description}

Providing contextual information is crucial for tailoring the test cases to the specific application at hand and preventing hallucinations~\cite{lewis2021retrievalaugmented}. However, it is equally important to ensure that crafting this description does not pose a greater challenge than generating the test cases themselves. To more effectively manage this trade-off and secure the precise transmission of relevant information, a template has been designed to guide the application description:

\setlength{\fboxsep}{10pt} 

\vspace{20pt}
\begin{center}
\fbox{
\begin{minipage}[l]{0.8\textwidth}
    \begin{itemize}[label={}]
        \item \textbf{What is the software name?}
    
        \begin{itemize}[label=\textbullet]
            \item A single line with the desired name to refer to the software
        \end{itemize}
    
        \item \textbf{What is the software's main purpose?}
        \begin{itemize}[label=\textbullet]
            \item A brief description with no more than 3 lines 
        \end{itemize}
        
        \item \textbf{What is the platform type?}
        \begin{itemize}[label=\textbullet]
            \item The platform of the software: Web, mobile, ... 
        \end{itemize}
        
        \item \textbf{Feature description}
        \begin{itemize}[label=\textbullet]
            \item An overview of each feature from the software
        \end{itemize}
    \end{itemize}
\end{minipage}
}
\end{center}

\vspace{20pt}

This is the outcome of empirical experiments conducted with different hypothetical software, evaluating the results through a qualitative analysis to identify any potential gaps or missing information. After some investigation it was observed that it was necessary to fill one template per feature in the software. The results found were way better than trying to describe all the feature on a single template file. So the software's name, purpose and platform type could be the same, but the feature description field had to be updated for each feature.

\vspace{5pt}

\subsubsection{Prompt Design}
In the prompt elaboration, throughout the tests and references about prompt engineering, the input for the model was separated into three parts, with each one complementing the other, reducing the complexity and granting better quality in the responses \cite{proptEngineering2023}. Each prompt is using the concept of Role Prompting \cite{rolePrompt2023}, which provides the model context in the way it should interpret the input, thus, contributing into more elaborated responses from the model.

In Requirements, the prompt is consisted of the role of the model (AI as a QA Engineer working on a project), the project template mentioned above with all information filled according to the under testing software, and an output structure for the model response. As cited before, the role of the model is to specify how the AI will interpret the project template, since the goal is to generate test cases, the model is designed to have a more technical approach to the subject, and for the output format, a standardization was necessary to ensure that the model response is in a pattern so it can be received as input for further prompts. 

In the prompt of Test Conditions, in addition to the template and the role of the model, this prompt uses the answer of last task, in a way that the model has time to “think", analyzing and improving the response in each task. At the end, the model should return the results in JSON format, containing the Requirements grouped by functional or non-functional followed by the Test Conditions of each one, providing a more viable structure for the next task to receive this as an input. 

At the Test Cases task, the prompt uses the responses of the last two tasks, the model role and an output format as a structured test case, containing title, preconditions, steps, expected results, test data and test classification. Once this is the last task, also was requested by AI to generate an output in markdown, helping the conversion to a PDF document, this conversion is a way to improve the visualization of the final answer of the model.

\subsection{Evaluation setup}

In order to comprehensively assess test case generation, a real-world application in active production, Da.tes~\cite{datescesar}, was deliberately chosen for study. Da.tes stands as a web platform designed to create opportunities connecting startups, investors and business in a agile and precise way, with the purpose to provide, expand and materialize the opportunities to transform organizations and people's lives.

Da.tes has a diverse spectrum of features and functionalities, ranging from recommendations and appropriated matches for the profile of each user to a data-driven ecosystem that facilitates and encourages new business, decision making and supporting the elaboration of strategies. This diversity allows the evaluation of the effectiveness of the test case generation process across various components and functionalities within a singular application.

To attest a test case quality according to \cite{samera2021} there are at least 30 different factors on literature review that could be used. These factors may be less complex like the size of a test case to more detailed criteria like branch coverage - The percentage of the number of branches exercised by the test cases divided by the total number of branches - Which requires code access of application under testing \cite{Istqb_2023}. So in order to evaluate the quality of AI generated test cases it was necessary to choose the metrics that would not require code access, historic test case execution data or number of found issues once these info were unavailable for Da.tes platform. 

All the quality factors chosen to evaluate the test cases had to be based on their documentation quality sub-factors like \textbf{Clarity}, \textbf{Correctness}, \textbf{Completeness}, \textbf{Consistency}, \textbf{Understandable}, \textbf{Self-contained} and \textbf{Specific}. Even thought it was only possible to use documentation quality factor to retrieve feedback about AI generated test cases, this metric is mentioned by \cite{samera2021b} on their quantitative study as the most common way to measure a test case quality, cited by 211 studies of 404 evaluated in total.

With the metrics to evaluate the generated test cases established it was necessary to create an evaluation process in order to help to answer this study queries. The chosen method was create 2 Google Forms with questions evaluating 10 AI generated test cases along with 10 test cases manually written by a QA team and share with 10 QA Engineers. The evaluators (volunteers) do not know which test case was created by human or AI.

On the first form the idea was that for each test case under analysis the volunteer would need to answer 7 questions, one for each documentation quality sub-factor. This method would generate 140 questions to be answered only on the first form by each volunteer, so it was necessary prioritize the most critical quality factors, that according to \cite{Lai_2017} were \textbf{Correctness}, \textbf{Completeness} and \textbf{Consistency}. So in order to reduce volunteers workload from 140 question on the first form to 60 questions but keeping results quality based on key factors only this three quality factors were analysed by this study. The volunteers could rate with scores between 1-5 which meant :

\begin{table}[ht!]
\begin{center}
\begin{tabular}{|c|c|}
    \hline
    Score & Meaning   \\
    \hline
    1     & Very Poor \\
    \hline
    2     & Poor      \\
    \hline
    3     & Regular   \\
    \hline
    4     & Good      \\
    \hline
    5     & Very Good \\
    \hline
\end{tabular}
\end{center}
\caption{Scores definition that was used by volunteers to rate the test cases}
\label{tbl:scores}
\end{table}

On the second form two similar test cases (focused on testing the exact same path on the software) were placed side by side. There were 10 questions to volunteers to choose test A or test B. Again none of them knew which test was generated by AI or by Humans and also after choose one test case there was a justification field where they were able to provide insights why test A or B was chosen. With these feedback that will be totally subjective and not focusing on quality factors as first form this study tries to unravel whether there are clear differences between both test cases group that deserves more attention. The whole process is described step by step on Table 2 below : 

\noindent
\begin{table}[ht!]
\begin{tabular}{p{0.1\textwidth}p{0.85\textwidth}}
\hline
\textbf{Steps} & \textbf{Description} \\ 
\hline \hline
Step 1&Generate test cases for da.tes website using AI. \\
\hline
Step 2&Get test cases written by da.tes QA team - Generated by humans. \\
\hline
Step 3&Make a Google Form listing 10 AI generated test cases and share with 10 QA Engineers volunteers. Asking them to assign scores from 1-5 regarding each documentation quality sub-factors. \\
\hline
Step 4&Make another Google Form with 10 AI generated test cases for da.tes along with 10 test cases written by da.tes QA team, and share with 10 QA Engineers volunteers. Select similar test cases (Test cases that have same objective, test same feature) and place them side by side in order to ask volunteers to rate which one is better. \\
\hline
Step 5&Analyse forms results and feedbacks in order to clarify assumptions and answer questions about the potential use of LLMs to generate test cases in a practical context. \\
\hline
\end{tabular}
\caption{Description of AI generated test cases evaluation process step by step}
\label{tbl:process_steps}
\end{table}

\vspace{20pt}

\section{Results and Discussions}


In total it was possible to get only 7 responses for both forms and this result analysis will be held on this small number of opinions. But the results between all the QA Engineers that answered the forms were quite similar, which tells that this study still could get relevant answers and insights. To avoid sampling bias it was calculated the average score for each group (AI and Human). While the IA generated tests average score was 4.31, the average result for humans was 4.18. Below it's explained the results summarized for each criteria.

Correction was the first quality factor analyzed and it held the most  discrepancy between the groups in the tests cases evaluation. Notably, the AI-generated test group exhibited better correctness evaluations when compared with the outcomes derived from manual testing.

The next quality factor evaluated was the consistency and although having similar results it was observed a higher level of consistency in the AI-generated tests groups. Consistency is defined as the degree of uniformity and reliability in test outcomes, reflects the ability of a set of test cases to produce reliable and expected results under varying conditions. With that in mind, its possible to assert that the test created by the AI maintain a certain level of logic similar to the tests developed manually.

Lastly, the remaining factor of quality analyzed was completeness. This aspect appeared to be the most similar quality factor across the test groups when considered in a comparative manner. Completeness, in the context of software testing, refers to the extent to which a set of test cases covers all possible scenarios and functionalities of the software. In this study, the completeness of test cases was found to exhibit similar characteristics between the AI-generated and manually generated groups. In the Figure \ref{image:average} below, its possible to visualize the average result of both test groups side by side. 

\begin{figure}[ht!]
\begin{center}
\includegraphics[width=11cm, height=7cm]{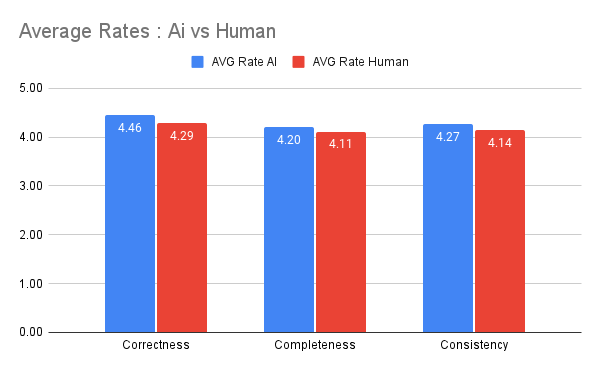}
\caption{Chart comparing the average rate for AI and Human test cases group split by quality factors}
\label{image:average}
\end{center}
\end{figure}

\begin{figure}[ht!]
\begin{center}
\includegraphics[width=11cm, height=7cm]{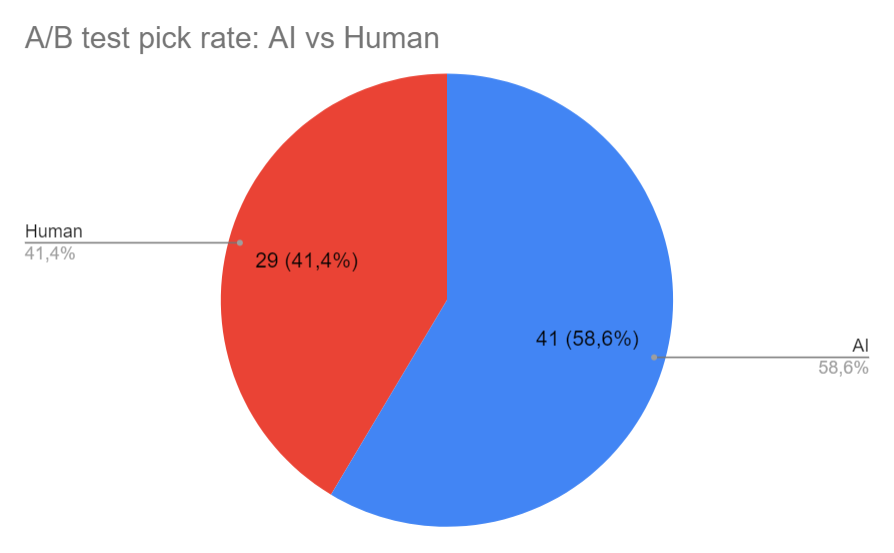}
\caption{Chart comparing the pick rate for AI and Human test cases in the A/B test}
\label{image:pizza_chart}
\end{center}
\end{figure}

The second questionnaire provided a more in-depth information about the differences between test cases designed with and without the involvement of generative AI. It was presented in a A/B testing format with similar tests grouped with the goal was to get the perceptions on the preferred test cases without inferring any specific metrics, and discover patterns. The results showed up that, out of the 10 test cases analyzed, 70 responses were obtained, with 41 answers pointing to the test cases generated by artificial intelligence (58.6\%\ of total). It is worth mentioning that in two specific test cases, there was a unanimous preference for human-developed test cases, while in one test case, the unanimous preference was for AI-generated test cases. These results are presented on Figure \ref{image:pizza_chart}.

When evaluating the characteristics attributed to test cases created by artificial intelligence, the quality of writing, simplicity, directiveness, clarity, and completeness stand out. These features resonated with a significant percentage of participants, suggesting a consistent appreciation of these elements in AI-generated test cases. This trend points to the effectiveness of artificial intelligence in producing tests that meet crucial criteria of quality and comprehensibility, as evidenced by the choices made by volunteers throughout the experiment.

\section{Conclusion}

This paper has presented a comprehensive case study investigating the integration of Large Language Models (LLMs) in the construction of test cases within the realm of software engineering. Through a systematic exploration employing a real software application (Da.tes), we have examined the practical efficacy of LLMs, aiming to illuminate their impact on test case construction processes.

The most relevant conclusion obtained by this study is that although there is a small numerical difference in the evaluations of the volunteers in this study that could suggest an advantage of the test cases developed through an AI, in general the average between the scores indicates that both groups obtained similar results to the point of stating that at least using AI (LLMs) to develop test cases resulted in artifacts of similar quality to those developed manually by a testing team.

It is worth mention that the process featured as “AI" is not completed automated since it depends on the human input to give the right format and information. The developed template was a build over many iterations and one of the findings is that was necessary to break the software feature by feature to achieve a non-generic results. 

As future directions -- since the evaluation results were similar between the two groups of tests -- it would be necessary to carry out additional studies comparing aspects such as Time, Cost and Learning Curve using LLMs to create test cases compared with manually creation. 

Also despite of the model work well with common features like login or sign up, the GPT3.5-Turbo could not handle many software features described on a single template, so it was observed that integration tests between two features could be a limitation of the current prompt design, once it requires that QA Engineer explicitly write the cross dependence between features, increasing the template size, but being cautiously to not write too much once the results could get worse as the feature description increases it size.

\bibliographystyle{apalike}
\bibliography{main}

\end{document}